# SADAS: an integrated software system for the data of the SuperAGILE experiment

Francesco Lazzarotto & Ettore Del Monte
CNR - Istituto di Astrofisica Spaziale e Fisica cosmica,
Roma, Via del Fosso del Cavaliere 100, 00133, Italy
https://www.linkedin.com/in/fralaz1971
mailto:fralaz1971@gmail.com

**Abstract**

**SuperAGILE (SA) is a detection system on board of the AGILE satellite (Astro-rivelatore Gamma a Immagini LEggero), a Gamma-ray astronomy mission approved by the Italian Space Agency (ASI) as first project for the Program for Small Scientific Missions, with launch planned in the second part of 2005. The developing and testing of the instrument took a big effort in software building and applications, we realized an integrated system to handle and to analyse measurement data since prototype tests until flight observations. The software system was created with an Object Oriented software design approach, and this permits to employ suitable libraries developed by other research teams and the integration of applications developed during our past work. This method allowed us to apply our schemas and written code on several prototypes, to share the work among different developers with the help of standard modeling instruments such as UML schemas. We also used SQL-based database techniques to access large amounts of data stored in the archives, this will improve the scientific return from space observations. All this has allowed our team to minimize the cost of developing in terms of man-power and resources, to dispone of a flexible system to face future needs of the mission and to invest it on other experiments.**

## 1. INTRODUCTION

The SuperAGILE experiment produces large amounts of data at high rate. The expected data stream is a continuous and massive flow (20 kb/s) of raw information sent to ground for a minimum of 3 years, plus a larger rate during ground tests, to handle them we have divided data processing in different steps. Data production, data preprocessing and archiving, data reduction and data final analysis. In this work we'll focalize on data reduction and analysis (final analysis = data reduction + data analysis). Data coming from the instrument concern physical measurements and equipment housekeepings. The main idea is to have a single application able to handle different operations from an unique console, then exploding it in sub-operation loops when requested and then returning to main control.

## 2. SA DATA ANALYSIS OVERVIEW

SA data are primarily organized in lists of events. Events are supposed to be generated by the interaction between physical observables and the detector, or by calibration pulses simulating the response of the instrument to high energy radiation. The base element for SA data are events carrying time, position and energy information for that event. The positional information for events is given by the address on the detector grid where the event is detected. We can refer to a detector sensitive unit (strip) with different coordinates systems (es. [strip, group]) suggested by the way the strips are related to some subcomponent of the instrument (Daisy Chains, detector,

SADAS: an integrated software system for SuperAGILE

**FIGURE 1:** Use cases for flight Super AGILE data

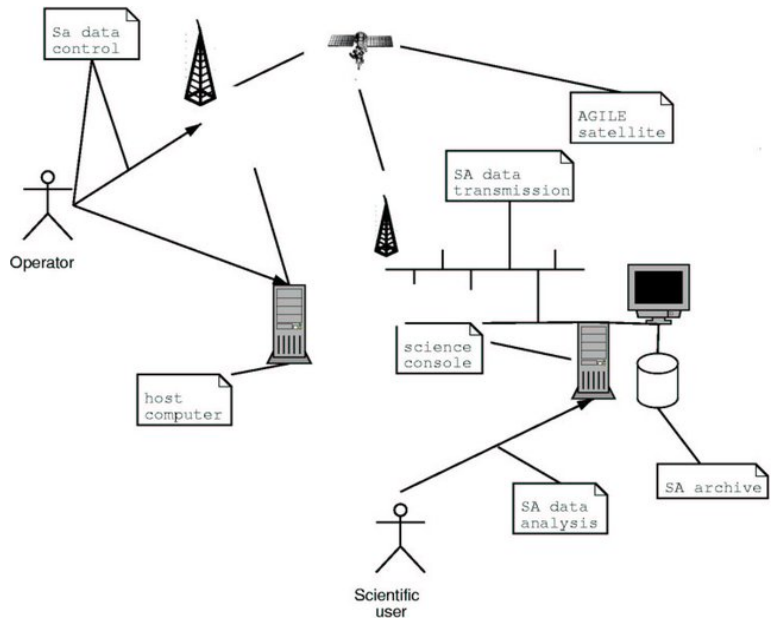

Chip, Group, ...). The positional information is given in output from the instrument as a sampled analogical measurement, this imply that address raw values are affected by measurement external conditions (mostly temperature but also radiation). To identify the right address output value we have to correct them using Look Up Tables created from calibration data, pulsing the addresses in a known succession and controlling that measured values in a fixed range match a determined address. SADAS is an integrated software for all kinds of scientific analysis on SA data, it's designed to accept SA data from different prototypes and to result suitable to analyse them. It's also designed to perform scientific analysis at different levels, from the data reduction, through the histogram products generation and visualization to statistical analysis and reports (e. g. best fit). Graphical User Interfaces (GUI) allow usability even to people non developers of the experiment.

**FIGURE 2:** Architecture of the SADAS software

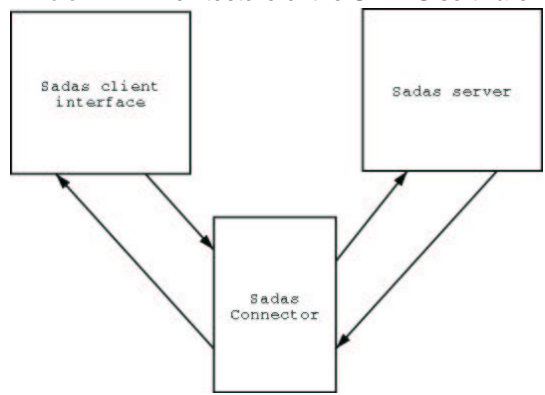

## 3. SADAS ANALYSIS LAYERS

The scientific analysis actions performed on SA data are organized in layers. This is to divide general operations from specific ones and to decouple some tasks in order to create an organized and optmized flow for the scientific operations. The first analysis layer (Prototype Layer) handles prototype dependent functionalities (SFTE, SITE, PDTE, FLHT). The second analysis layer (Dataset Layer) regards the subject of the read data and provides the application to perform only the operations possible on the selected kind of data. The third layer (Analysis Type Layer) branches to a subset of analysis operations related to a certain scientific analysis argument. Possible analysis types are Quicklook, Imaging, Timing, Energy, Calibration, Setup and Simulation. The fourth analysis layer allows to execute in the fixed analysis scope the basic steps of the analysis: Read, Process & Plot, Fit & Report. Every analysis step performed by the server is started receiving inputs from the user interacting with the GUI forms of the client we show in figure 3 on page 3 and in figure 4 on page 4.

**FIGURE 3:** SADAS main window

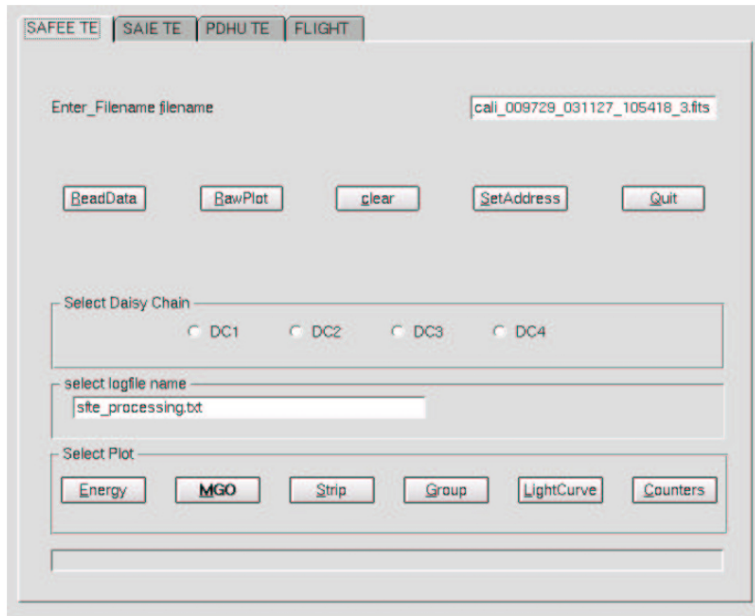

## 4. SADAS ARCHITECTURE

Sadas software is organized with a client-server architecture based also on a class (SadasConnector) which implements the communication channel abstraction (figure 2 on page 2).

**4.1. SADAS client**

Provides 4 (or more) layers of input instructions sets to pass to the server.

>The implementation can be changed without change the rest of the project
>Can run on the scientfic user local console
>implementations :
>>– C++, Trolltech Qt based GUI
>>– Command line IDL program

**4.2. Connection module**
>Transmits sending and receiving instructions (messages) between the client and the Sadas scientific operations server
>Different connection types have the same calling interface callable by the client and by the server
>implementations:
>>– Communication temporal files
>>– Network sockets for remote analysis connections
>>– Shared memories

### 4.3. SADAS server

Porvides 4 (or more) layers of operations in the analysis flow, the implementation can be changed without changing the rest of the project. It's an independent application, separated from the client user interface and can run on the data archive server, even in a remote location respect to end-line users. Scientific operations are performed using Object Oriented libraries based on Root CERN package and RSI-IDL Astrolib.

**FIGURE 4:** SADAS energy analysis console

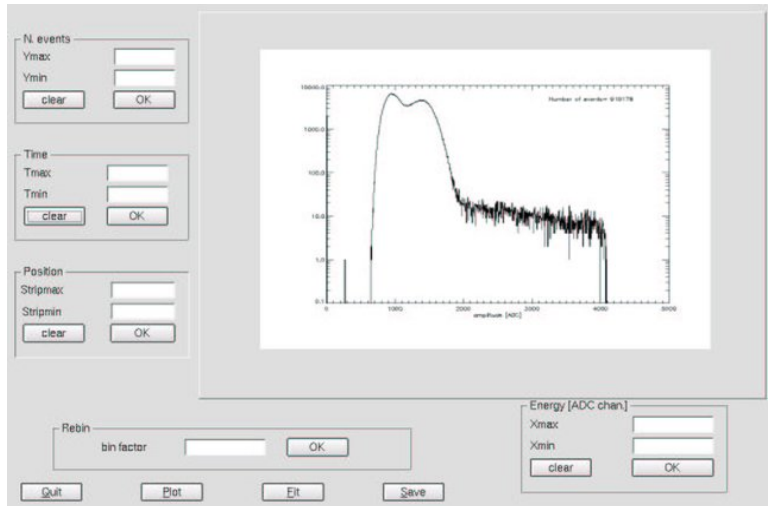

### 5. CONCLUSIONS

A modern scientific experiment produces huge amounts of different kinds of data, an experimental team needs more specific competence on hardware and software developing. Modern Computer Science techniques in design, developing strategies, architecture and algorithms are now widely appliable with good enhancement on scientific return. Even for a small mission like AGILE with few economical resources to get manpower and instruments is possible to produce high level software in terms of usability and processing performances, this is possible using public domain open source libraries and tools, well tested by scientific community, developing the applications with the philosophy of "evolved products able to evolve".

### 6. FUTURE WORK

We plan to use this experience to ultimate a complete scientific analysis tool to apply on SA data produced after launch planned for July 2005. Will be easier and easier to extract methods and a reusable library to apply the work done to future scientific experiments. An important final enhancement could be to port the system in an omogeneous software platform suitable for all requested tasks for a real time scientific experiment, we think to Java but it's not a definitive conclusion (see [1] and [2]). As a matter of fact it provides features as to be multipurpose, a clear developing interface, to be totally Operating System independent and to be network applications dedicated.